\documentclass{svjour3}
\smartqed  
\usepackage{graphicx}
\usepackage{dsfont}
\newcommand{\be}{\begin{equation}}
\newcommand{\ee}{\end{equation}}

\begin{document}                                                                                   

\noindent
{\Large \bf Fermionic boundary conditions and the finite 
\vskip1mm
\noindent
temperature
transition of QCD } 

\vspace{12mm}
\noindent
{\bf Erek Bilgici$\,^a$, Falk Bruckmann$\,^{b,c}$, Julia Danzer$\,^a$,
Christof Gattringer$\,^a$, Christian Hagen$\,^c$,
 Ernst Michael Ilgenfritz$\,^d$, Axel Maas$\,^a$}

\vspace{5mm}
\noindent
$^a$Institut f.~Physik, Universit\"at Graz, 8010 Graz, Austria\\
$^b$Institut f.~Theoretische Physik III, Universit\"at Erlangen, 91058 Erlangen, Germany\\
$^c$Institut f.~Theoretische~Physik, Universit\"at Regensburg, 93040 Regensburg, Germany\\
$^d$Institut f.~Theoretische Physik, Universit\"at Heidelberg, 69120 Heidelberg, Germany

\vspace{20mm}

\begin{abstract}
$ $
\\
\\
Finite temperature lattice QCD is probed by varying the temporal boundary
conditions of the fermions. We develop the emerging physical behavior in a study
of the quenched case and subsequently present first results for a fully
dynamical calculation comparing ensembles below and above the phase transition.
We show that for low temperature spectral quantities of the Dirac operator are
insensitive to boundary conditions, while in the deconfined phase a non-trivial
response to a variation of the boundary conditions sets in.
\footnote{Contribution prepared for a special issue of  {\sl Few Body Systems}
on the occasion of Willibald Plessas' sixtieth  birthday. 
Happy birthday Willi !} 
\end{abstract}

\vspace{8mm}
 
\section{Introduction}
Understanding the nature and mechanisms of the QCD phase transition  has
recently  become one of the great issues of non-perturbative QCD studies.
Questions such as {\sl "How can one characterize the high temperature plasma
phase?"}, or {\sl "What are the field excitations that drive the transition?"},
are still far from having a generally accepted answer. A powerful
approach to analyzing such questions is the formulation of QCD on a
Euclidean space time lattice. In this setting Monte Carlo simulations allow one
to obtain non-perturbative results from an ab-initio calculation.   

In the Euclidean formulation one dimension, the Euclidean time, is compactified,
turning the base manifold into a (hyper) cylinder. The circumference of the
cylinder is the inverse temperature. Thus increasing the temperature means
shrinking the temporal extent of the lattice. If the temperature is sufficiently
high, the temporal extent is shorter than the relevant scale $\Lambda_{QCD}$,
and correlations around the compact time direction change the physics. The onset
of such a strong self-correlation around time corresponds to the critical
transition temperature $T_c$.   

This qualitative insight about the relation of the scale $\Lambda_{QCD}$
and the time extent given by the inverse temperature can be tested on the
lattice. A powerful approach to such an analysis is the use of the temporal
boundary conditions as a tool. Instead of using the canonical choice -- periodic
boundary conditions for the gauge fields and anti-periodic boundary conditions
for the fermions -- one may implement more general temporal boundary conditions
to probe the system. According to the qualitative picture outlined above, one
is inclined to expect that below $T_c$  the system remains relatively
unchanged under varying boundary
conditions, while above $T_c$ the response of the system to a change
of the boundary conditions should be strong.

In a series of papers the strategy of probing finite temperature QCD with the
help of boundary conditions has been explored in quenched calculations. In
particular generalized fermion boundary conditions have been considered. Various
spectral quantities of the Dirac operator were studied 
\cite{GaSch,Ga1,Jena1,Jena2,kovacs,BiGa,soeldner}. 
A new observable was proposed and studied \cite{dualcond1,dualcond2,erek,fischer}, the dual
chiral condensate, which probes the dependence on the boundary conditions and
provides a link between the Polyakov loop and the  conventional chiral
condensate. 

The analysis was also extended beyond the gauge group SU(3). Quenched studies
for the center-trivial group G$_2$  \cite{G2} and for SU(2) gauge theory with
adjoint fermions \cite{Su2adj} were conducted.

In all these quenched studies it was established, that below $T_c$ the Dirac
spectrum is insensitive to the temporal fermion boundary conditions.  Above the
deconfinement transition, which in the quenched case is characterized by the
emergence of a non-vanishing expectation value of the Polyakov loop, a 
non-trivial response to changing temporal boundary conditions sets in for 
fermionic quantities such as the chiral condensate and the spectral gap. It was
furthermore observed \cite{GaSch,dualcond1,dualcond2,erek,Z3paper,Luschevskaya} that only
the relative phase between the phase of the Polyakov loop and the phase in the
fermionic boundary condition is relevant for the physics. This even holds for
the center-trivial group G$_2$, which behaves similar to SU(N) when one
restricts the high temperature ensembles to the center sector characterized by a
real Polyakov loop. 

Interesting is also the case of SU(2) (more generally SU(N)) with adjoint
fermions \cite{Su2adj}, a theory where the deconfinement and chiral symmetry
restoration temperatures do not coincide, the latter being considerably higher
than the former. In agreement with the picture outlined above, at the
deconfinement temperature the spectrum becomes sensitive to the boundary
conditions. The intermediate phase between the deconfinement and chiral
restoration temperatures is characterized by a non-vanishing density of
eigenvalues near the origin, and thus, due to the Banks-Casher relation
\cite{BaCa} by a non-vanishing chiral condensate, which, however, varies in size
with the boundary condition. Only above the chiral restoration temperature the
condensate vanishes completely for the physical anti-periodic fermion boundary
conditions.  

In the first part of this contribution, in Section 2, we discuss the physical
situation for the interplay of boundary conditions and spectral observables for
the quenched case. In the second  part, in Section 3, we present first
preliminary results for lattice QCD with dynamical fermions. In particular we
analyze publicly available dynamical SU(3) configurations generated for two
flavors of staggered sea quarks by the MILC collaboration \cite{milc}. Ensembles
below and above the QCD transition are studied with the staggered Dirac operator
and we compare the response of various spectral quantities below and above $T_c$
to changing boundary conditions. In Section 4 we summarize our  results and give
an outlook.       

\hfill

\section{The physical situation for the quenched case}
 
As outlined in the introduction, the interplay of fermionic boundary conditions
and the spectral quantities of the Dirac operator below and above $T_c$ so far
was analyzed only for the quenched case. Although by neglecting the dynamical
quark effects the quenched case certainly does not reflect the full problem, as
a toy model it has one important conceptual advantage: It represents the 
passive response of the (fermionic) observables for a sudden (smooth for SU(2))
transition of the gluonic environment. For SU(N), N $> 2$, there is a first
order transition between the two  phases and one can expect to find a distinctly
different behavior of the  Dirac operator below and above $T_c$. The fully
dynamical theory, on the other hand, shows only a crossover, thus  leading to a
gradual variation of the properties rather than a sharp change. For that reason
the quenched case is an interesting model laboratory.

The calculations were done on SU(3) gauge ensembles with 100 configurations
each, generated with the L\"uscher Weisz gauge action \cite{LuWe} with tadpole
improvement. When we quote
results in physical units, these were obtained from a determination of the scale
\cite{scale} using the Sommer parameter. Our fermionic observables were computed
for the staggered Dirac operator (we here set the lattice spacing to $a = 1$)
\begin{equation}
D(x,y) \; = \; \sum_\mu \, \eta_\mu(x) \, 
\Big[ \; U_\mu(x) \, \delta_{x+\hat{\mu},y} 
\, - \, U_\mu(x-\hat{\mu})^\dagger \, \delta_{x-\hat{\mu},y} \; \Big] \; ,
\label{diracop}
\end{equation}
where $\eta_\mu(x)$ is the staggered sign function $\eta_\mu(x) = 
\prod_{\nu=1}^{\mu-1}(-1)^{x_\nu}$.  For the massless staggered lattice Dirac
operator (\ref{diracop}) we evaluate  complete eigenvalue spectra using a
parallel implementation of standard linear algebra routines. The staggered Dirac
operator is anti-hermitian and consequently the eigenvalues $\lambda^{(j)}$ are
purely imaginary. The eigenvalues for the Dirac operator with mass $m$ are then
given by $\lambda^{(j)} + m$. From the complete eigenvalue spectra all our
fermionic observables may be computed. 

As outlined in the introduction, we probe the system by changing the temporal
fermionic boundary conditions. These are introduced as,
\begin{equation}
\psi(\vec{x},N_T) \; = \; e^{i\phi} \, \psi(\vec{x},0) \; ,
\label{fermionbc}
\end{equation}
where $N_T$ is the number of lattice points in the compact time direction and
the "boundary angle" $\phi \in [0,2\pi]$ parameterizes the boundary conditions.
Bilinear combinations (which are bosons) are periodic.
A value of $\phi = \pi$ corresponds to the usual anti-periodic boundary
conditions. Other values are adopted to analyze the system. In particular we
consider a total of eight (for some tests also 16 up to 128) 
equally spaced intermediate values of $\phi$ in the
interval $[0,2\pi]$. For completeness we stress, that all other boundary
conditions, i.e., the spatial fermionic boundary conditions and the boundary
conditions for the gauge fields, were kept periodic. We work with various
lattice sizes ranging from $8^3\times 4$ to $14^3 \times 6$, with typically 100 
configurations per ensemble. All errors we show are statistical errors
determined with single elimination Jackknife.

\begin{figure}[t]
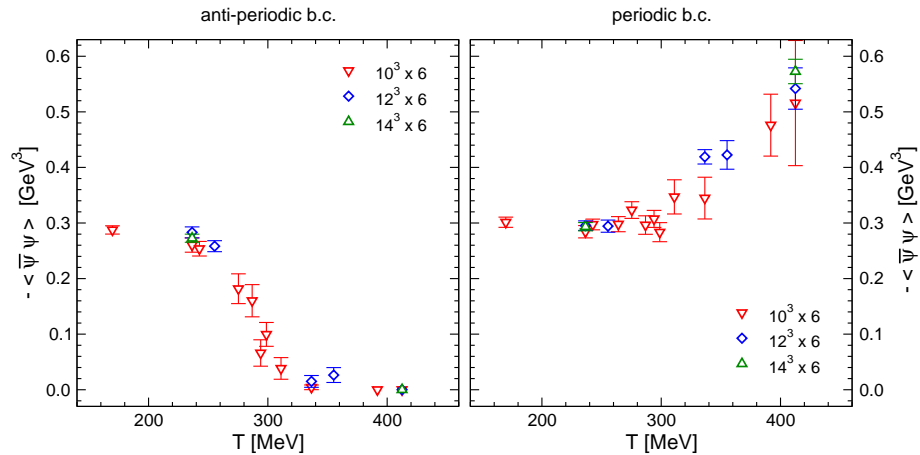

\begin{center}
\includegraphics*[width=6cm]{cond_apbc.eps}
\includegraphics*[width=6cm]{cond_pbc.eps}
\end{center}
\caption{\label{fig:cond_quench}
The bare quenched chiral condensate $-\langle \overline{\psi} \psi \rangle$ in
GeV$^3$ for anti-periodic (lhs.\ plot) and periodic (rhs.) temporal fermion
boundary conditions as a function of temperature.}
\end{figure}

Let us begin the discussion of observables with the bare (unrenormalized)
chiral condensate. In terms of the eigenvalues it is given by 
\begin{equation}
- \, \langle \overline{\psi} \psi \rangle \; = \; 
\lim_{m \rightarrow 0} \lim_{V \rightarrow \infty} \, \frac{1}{V} \, 
\mbox{Tr} \, (D + m)^{-1} \; = \;
\lim_{m \rightarrow 0} \lim_{V \rightarrow \infty} 
\frac{1}{V} \sum_j \frac{1}{\lambda^{(j)} + m} \; ,
\label{condensatesum}
\end{equation}
where $V$ is the 4-volume and $m$ the quark mass parameter. Obviously it is not
possible to perform the thermodynamic limit on a finite lattice. Two alternative
approaches are possible: Firstly, one may analyze the condensate for finite
volume as a function of the mass parameter $m$ and extrapolate to $m = 0$,
ignoring the sharp drop to 0 at $m = 0$, which must appear as long as the volume
is finite. Secondly, one can use the Banks-Casher formula \cite{BaCa}, which
relates the chiral condensate to the spectral density  $\rho(0)$ at the origin, 
\begin{equation}
- \, \langle \overline{\psi} \psi \rangle \; = \; \pi \, \rho(0) \; ,
\label{bacarel}
\end{equation}
and determine $\rho(0)$ to compute the chiral condensate. It can be shown
\cite{G2} that for a fixed volume both methods give rise to perfectly consistent results.

In Fig.~\ref{fig:cond_quench} we show the bare chiral condensate $-\langle
\overline{\psi} \psi \rangle$ as a function of the temperature. In the lhs.\
plot we display the results for the physical anti-periodic temporal fermion
boundary conditions, while the rhs.\ plot is  for periodic boundary conditions.
It is obvious, that the two cases behave rather differently. For the physical
anti-periodic boundary conditions the condensate starts to melt near the
critical temperature of $T_c \simeq 280$ MeV, while for the periodic case the
bare condensate even rises beyond $T_c$.  Below the critical temperature the
anti-periodic and the periodic data essentially agree, thus reinforcing the
expectation that below $T_c$ the system is insensitive to the imposed boundary
conditions, while above $T_c$ (many) physical quantities show a non-trivial
dependence on the boundary conditions. 

\begin{figure}[t]
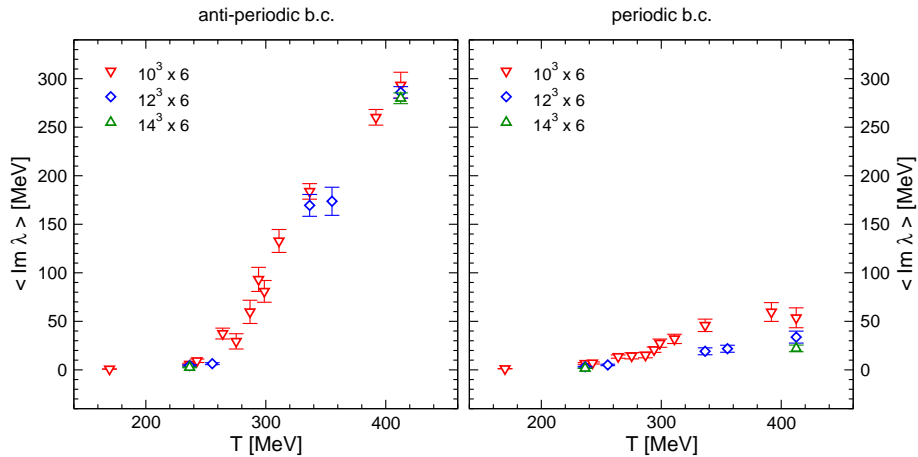

\begin{center}
\includegraphics*[width=6cm]{gap_apbc.eps}
\includegraphics*[width=6cm]{gap_pbc.eps}
\end{center}
\caption{\label{fig:gap_quench}
The quenched spectral gap in MeV for anti-periodic (lhs.\ plot) and periodic
(rhs.) temporal fermion boundary conditions as a function of temperature.}
\end{figure}

According to the Banks-Casher relation \cite{BaCa} a vanishing condensate, as we
observe it for the anti-periodic boundary conditions at high temperatures, must 
coincide with a vanishing spectral density $\rho(0)$. Such a vanishing density 
above $T_c$  is usually attributed to the opening up of a gap in the spectrum.
We analyze the spectral gap using the expectation value $\langle |
\lambda_{min} | \rangle$, where $\lambda_{min}$ is the smallest eigenvalue of
the Dirac operator.

In Fig.~\ref{fig:gap_quench} we show the results for the spectral gap as a
function of the temperature, where we again compare the physical anti-periodic
boundary conditions (lhs.) to the periodic case (rhs.). As for the condensate,
we find also for the spectral gap that up to $T_c \simeq 280$ MeV the two cases
give roughly the same results, while above $T_c$ we observe a non-trivial
response to a change of the boundary conditions. For the anti-periodic boundary
conditions, we find that above $T_c$ the spectral gap opens up quickly in
accordance with the vanishing condensate for that case demonstrated in the lhs.\
plot of Fig.~\ref{fig:cond_quench}. When inspecting the periodic case on the
rhs.\ plot of Fig.~\ref{fig:gap_quench} we do not observe the opening up of a
spectral gap. There is a slight upward trend, but way less pronounced than for
the anti-periodic boundary conditions. This upward trend for the periodic case
also becomes weaker as the spatial volume is increased and thus is probably a
finite size effect, similar to the "microscopic gap" $\propto 1/V_{space}$ known
from random matrix theory \cite{randommatrix}. 

Although the picture for the spectral gap seems rather clear, there is an
interesting alternative scenario to be considered: The vanishing spectral
density at the origin, $\rho(0)$, which according to the Banks-Casher formula (\ref{bacarel})
is necessary for a vanishing chiral condensate, does not necessarily imply a
spectral gap. An alternative would be a spectral density that is non-vanishing
for all non-zero eigenvalues but continuously goes to zero at the origin. This
would imply that the spectral gap of the lattice Dirac operator, shown in the
lhs.\ plot of  Fig.~\ref{fig:gap_quench}, closes as the spatial volume is sent
to infinity. The currently available data do not allow one to decide between a
closing gap at infinite spatial volume or a finite limit for the gap
\cite{kovacs,kovacs2}. 

In this context we also stress that the staggered lattice Dirac operator is
numerically cheap, but certainly not the first choice for analyzing the chiral
condensate and the spectral gap. The reason is that the staggered Dirac operator
does not discriminate between low lying eigenvalues and eigenvalues that
correspond to zero modes originating from topological excitations. The latter,
however, should be removed in both the determination of the spectral gap, as
well as for the evaluation of the spectral density $\rho(0)$. Using, e.g., the
overlap operator would allow for a cleaner study -- at a considerably higher
computational cost, however.

We have seen that above $T_c$ the spectral gap and the chiral condensate, which
in turn through (\ref{bacarel}) is related to the spectral density $\rho(0)$ at
the origin, are sensitive to the fermionic boundary conditions. Both the
spectral gap and the density $\rho(0)$ are infrared properties of the Dirac
spectrum. An interesting question is how the other parts of the Dirac spectrum
respond to changing boundary conditions. This question was analyzed numerically
in \cite{dualcond,erek} and analytically in \cite{Jena2}. It was found that the
IR modes of the Dirac spectrum are most sensitive to a change of the boundary
conditions, while the shift of the eigenvalues under a variation of the boundary
angle $\phi$ decreases exponentially as one moves towards the UV end of the
spectrum. 

\begin{figure}[t]
\begin{center}
\includegraphics*[width=8cm]{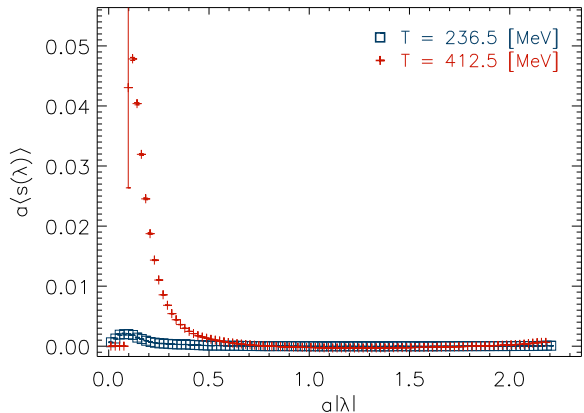}
\end{center}
\caption{\label{fig:evalshift_quench}
The expectation value of the shift variable $s(\lambda)$ as a function of
$|\lambda|$ (both in lattice units). The data are for the quenched $14^3 \times
6$ lattices and we compare two ensembles below ($T = 236.5$ MeV) and above $T_c$
($T = 412,5$ MeV).}
\end{figure}

In order to quantify this shift, in Fig.~\ref{fig:evalshift_quench} we show the
shift $s(\lambda)$ of the eigenvalues when comparing 
boundary conditions with $\phi = \pi/2$ and boundary conditions  with
boundary angle $\phi = \pi$. More explicitly, the shift variable $s(\lambda)$
is defined as
\begin{equation}
s(\lambda) \; = \; \mbox{Im} \, 
\Big( \lambda_{\phi = \pi/2} \, - \, \lambda_{\phi = \pi} \big) \; .
\label{shiftvariable}
\end{equation}
In Fig.~\ref{fig:evalshift_quench} we plot this shift observable as a function
of the size of the physical ($\phi = \pi$) eigenvalue (both quantities in lattice units), and compare 
the results below and above $T_c$. It is obvious that in both cases a
noticeable shift is seen only for the IR part of the spectrum, and $s(\lambda)$
decreases quickly towards the UV end of the spectrum. The plot also again
confirms that only above $T_c$ there is a sizable shift of the eigenvalues when
changing the boundary conditions. Let us finally comment on the drop of the
shift variable at $\lambda \simeq 0$: We attribute this effect to the would-be
zero modes of the staggered Dirac operator. For a chiral operator such as the
overlap operator, they would be frozen at the origin when changing the boundary
conditions. The staggered operator does not protect them from a shift, but the
nature of the would-be zero modes is at least manifest in a drop of the shift
variable $s(\lambda)$ at $\lambda \simeq 0$.

\begin{figure}[t] \begin{center} 
\includegraphics*[width=8cm]{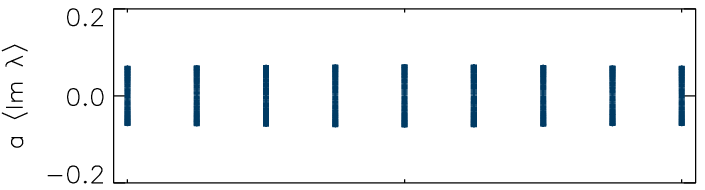}
\vskip2mm
\includegraphics*[width=8cm]{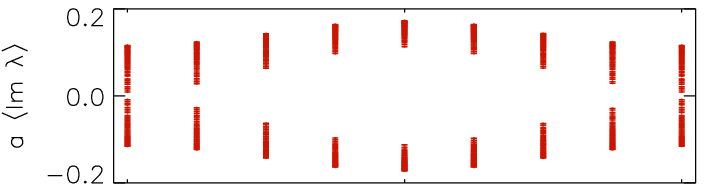}
\vskip2mm
\includegraphics*[width=8cm]{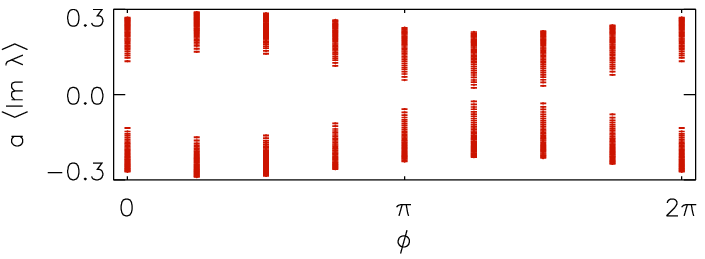}
\end{center} 
\caption{\label{fig:loweval_quench}   
We show the lowest 40 eigenvalues of the lattice Dirac operator for various
values of the boundary angle $\phi$ using our $14^3 \times 6$ quenched ensembles
($8^3 \times 4$ for the bottom plot). The top plot is for a temperature below
$T_c$, while the other two plots are for $T>T_c$. They differ by the sector of
the Polyakov loop: In the center plot we use a configuration in the real
Polyakov loop sector, while at the bottom the eigenvalues are shown for the
sector where the phase of the Polyakov loop is close to $2\pi/3$.} 
\end{figure}

So far we have only compared two values of the boundary angle $\phi$. In
Fig.~\ref{fig:loweval_quench} we now show how the lowest 40 eigenvalues of the
Dirac operator in the quenched case behave as a function of $\phi$. It is
obvious, that for ensembles below $T_c$ (top plot) the eigenvalues are
essentially independent of $\phi$. Above $T_c$ they show a sine-like behavior as
a function of $\phi$. For the sector where the Polyakov loop is essentially real
(center plot), the lowest eigenvalues come close to zero for the case of
periodic boundary conditions ($\phi = 0$).  For the case of configurations in a
sector with complex Polyakov loop (bottom plot), the displacement pattern of the
eigenvalues is shifted by $\pm 2\pi/3$.

For the quenched case the Polyakov loop is an order parameter for the
deconfinement transition \cite{Polyakov}. It signals the breaking of the center
symmetry, which for the quenched theory is manifest below $T_c$ and the
expectation value of the Polyakov  loop is zero then. Above $T_c$ the center
symmetry becomes broken spontaneously and the Polyakov loop acquires a
non-vanishing expectation value. Its phase spontaneously selects one of the
three (for SU(3)) values, $0, 2\pi/3, - 2\pi/3$. 

One of the original motivations \cite{Ga1} for analyzing the connection between
boundary conditions and the Dirac spectrum, was to find out, how the change of
the Polyakov loop at the phase transition affects properties of spectral sums of
Dirac eigenvalues. The Polyakov loop is an order  parameter for confinement,
while certain spectral sums of the Dirac operator,  e.g., the one in
Eq.~(\ref{condensatesum}), are related to chiral symmetry. Thus understanding
the relation between the Polyakov loop and the Dirac spectrum might provide an
understanding of a possible relation between chiral symmetry breaking and
deconfinement. It is natural to expect that the Dirac operator and its spectral
properties connect both, confinement and chiral symmetry. After all,
the quark-propagator, i.e., the inverse Dirac  operator, should know whether the
quark it describes is in the chirally  symmetric or the chirally broken phase,
and if it is confined or not.  

For understanding the relation between the chiral condensate and the 
Polyakov loop we write the condensate in a rather general form: The chiral 
condensate is a gauge invariant quantity, and as such may be written as a
sum of (traced) gauge transporters along closed loops on the lattice,
\begin{equation}
- \overline{\psi} \psi \; = \; \sum_{l \in {\cal L}} c(l) \; \mbox{Tr} \!\!\!
\prod_{(x,\mu) \in l} U_\mu(x) \; .
\label{loopexpansion}
\end{equation}
Here ${\cal L}$ is the set of all loops that contribute, $l$ is an individual
loop in this set, $c(l)$ a complex valued coefficient, and the product runs
over all links in $l$. We remark that the structure in Eq.~(\ref{loopexpansion})
is universal and different lattice discretizations of the lattice Dirac operator only lead to different values for the coefficients $c(l)$. 

On a finite lattice it is possible to order all loops $l$ in (\ref{loopexpansion})
according to their winding number $q$ around the compact time direction. If we implement the boundary condition (\ref{fermionbc}), then the 
loops pick up a phase $\exp(i\phi)$ with every winding. Thus we can write 
the condensate for the general boundary condition (\ref{fermionbc}) as
\begin{equation}
- \overline{\psi} \psi \, \Big|_\phi \; = \; \sum_{q \in \mathds{Z}} 
e^{i \phi q} \sum_{l \in {\cal L}^{(q)}} c(l) \; \mbox{Tr} \!\!\!
\prod_{(x,\mu) \in l} U_\mu(x) \; ,
\end{equation}
where ${\cal L}^{(q)}$ is the set of all loops that wind exactly $q$-times. 
Using a Fourier transformation with respect to the boundary angle we can 
project to the contributions that wind exactly once, and in this way define 
the dual chiral condensate $\Sigma^{(1)}$,
\begin{equation}
\Sigma^{(1)} \; = \; - \int_0^{2\pi} \frac{d\phi}{2\pi} \,
\overline{\psi} \psi 
\Big|_\phi \, e^{- i \phi} \; = \; 
\sum_{l \in {\cal L}^{(1)}} c(l) \; \mbox{Tr} \!\!\!
\prod_{(x,\mu) \in l} U_\mu(x) \; ,
\label{dualconddef1}
\end{equation}
where we used $\int d\phi \exp(i \phi(q-1)) = 2\pi \delta_{q,1}$ in the second
step. The dual chiral condensate $\Sigma^{(1)}$ consists of the loops of the
conventional chiral condensate, but restricted to winding number 1. These loops,
however, transform under a center transformation in exactly the same way as the
Polyakov loop. Consequently, $\Sigma^{(1)}$ also serves as an order parameter
for the breaking of the center symmetry and thus in the  quenched case is an
order parameter for confinement. We remark that the technique of using a Fourier transformation with respect to the boundary angle has been used in various contexts, in particular for the construction of canonical determinants, i.e., fermion determinants that describe a fixed quark number.

We stress at this point, that we have not performed the limits $V \rightarrow
\infty$ and  $m \rightarrow 0$ in the definition (\ref{dualconddef1}) of the
dual chiral condensate,  but of course these two limits are necessary when one
wants to use  $\Sigma^{(1)}$ for the analysis of chiral symmetry. On the other
hand, since the weight factors $c(l)$ in (\ref{dualconddef1}) behave for large 
$m$ as $c(l) \propto m^{-|l|}$, where $|l|$ is the length of the loop $l$,  the
infinite mass limit reduces the dual condensate $\Sigma^{(1)}$ to the shortest
loops that wind exactly once, i.e., the conventional straight  Polyakov loops. 

For a practical evaluation of the dual condensate, we make use of the spectral 
representation (\ref{condensatesum}) and obtain 
\begin{equation}
\Sigma^{(1)} \; = \; \int_0^{2\pi} \frac{d\phi}{2\pi} \,
S(\phi)
\, e^{- i \phi} \; , \qquad \mbox{with} \quad S(\phi) \; = \; 
\frac{1}{V} \sum_j \frac{1}{\lambda^{(j)}_\phi + m} \; ,
\label{dualconddef2}
\end{equation}
where $\lambda^{(j)}_\phi$ denotes the $j$-th eigenvalue computed for boundary
angle $\phi$. The $\phi$-integral is evaluated numerically using the eight
intermediate values of $\phi$ in the interval $[0,2\pi]$. Increasing the number
of intermediate values further leads only to corrections smaller than one
percent. 

\begin{figure}[t]
\begin{center}
\includegraphics*[width=7cm]{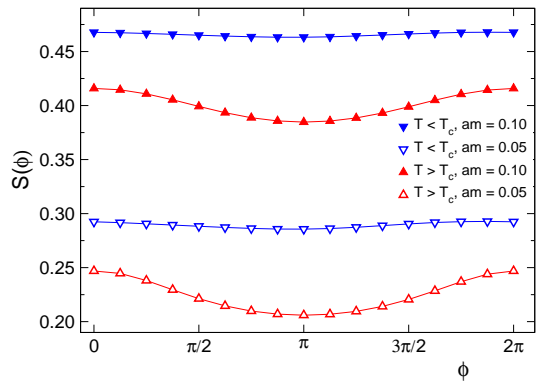}
\end{center}
\caption{\label{fig:integrand_quench}
The integrand $S(\phi)$ of the dual chiral condensate in the quenched case as a
function of the boundary angle $\phi$. We show the results for two bare quark
masses and compare ensembles below and above $T_c$ 
(Figure from \cite{dualcond1}.).}
\end{figure}

The dual chiral condensate (\ref{dualconddef2}) is the first  Fourier component
of the spectral sum $S(\phi)$. For small quark mass $m$ this spectral sum is
obviously dominated by the IR end of the spectrum and  thus should show a
non-trivial dependence on $\phi$ above $T_c$, where the  IR eigenvalues move
when changing boundary condition. Below $T_c$ the eigenvalues are essentially
independent of $\phi$ and thus $S(\phi)$ is approximately  constant, and
(\ref{dualconddef2}) implies a vanishing dual chiral condensate  below $T_c$.
This behavior of the spectral sum $S(\phi)$ is obvious from 
Fig.~\ref{fig:integrand_quench}, where we show $S(\phi)$ as a function of 
$\phi$ at two different bare quark masses $m$ for two ensembles below and above
$T_c$. 

\begin{figure}[t!]
\begin{center}
\includegraphics*[width=9cm]{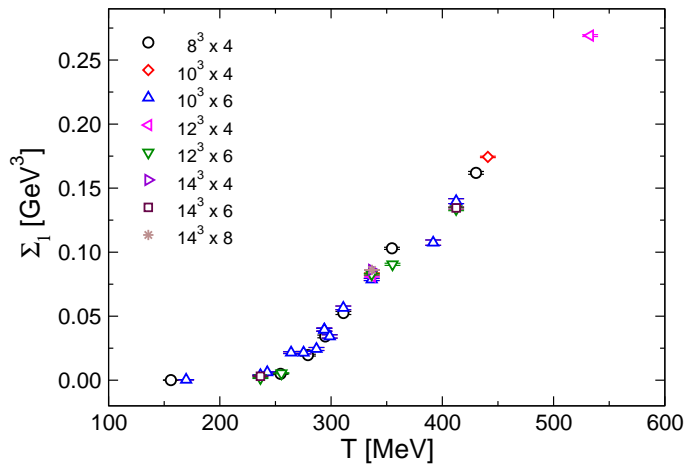}
\end{center}
\caption{\label{fig:dualcond_quench}
The quenched dual chiral condensate in physical units as 
a function of the temperature (Figure from \cite{dualcond1}.).}
\end{figure}

Finally in Fig.~\ref{fig:dualcond_quench} we present the (unrenormalized) dual
chiral condensate $\Sigma^{(1)}$ as a function of the temperature $T$. We show
results for  different lattice volumes and lattice spacings. The bare quark mass
was  chosen to be $m$ = 100 MeV for all these ensembles. Obviously the dual chiral 
condensate vanishes below $T_c \simeq 280$ MeV, while above the transition  it
acquires a non-vanishing value which rises quickly with $T$. It is remarkable
that the data points from the rather different ensembles  fall essentially on a
universal curve, which cannot a-priori be expected for an unrenormalized
quantity (the conventional thin Polyakov loop is a counterexample). This
hints at simple renormalization properties of the dual chiral condensate.  

\section{Results for QCD with dynamical fermions}

We now come to
discussing the results for QCD with dynamical fermions. Several aspects change
when turning on the sea quarks. There is no longer a sharp transition, but a
continuous crossover. Consequently we can expect only a gradually changing
behavior of the system. Also the center symmetry is broken explicitly by the
fermion determinant. Thus the Polyakov loop and also the dual chiral condensate
have a non-vanishing expectation value at low temperatures, which predominantly
comes from contributions of the sector with quark number $q = -1$.  

Although several aspects are different for the dynamical case, one can ask 
the question how much of the physical properties which we discussed for the 
quenched case are manifest also in the full theory. We will show, that 
indeed the qualitative behavior is unchanged, with a small sensitivity to 
the temporal fermion boundary conditions at low temperatures, and a strong 
response to a variation of the boundary angle $\phi$ in the high temperature 
regime. 

The dynamical configurations we analyze are from the publicly available 
ensembles for two flavors of staggered fermions, provided by the {\sl"Gauge
Connection"} \cite{milc}. In particular  we consider the $12^3  \times 4$
lattices, and focus on the configurations  generated with a bare quark mass of
$m = 0.008$ in lattice units. They are available for six temperature values
in a small band near $T_c \simeq 153$ MeV, with temperatures $T = 149.8,
151.5, 153.2, 155.0, 156.8$ and $160.4$ MeV, according to the scale determined
in \cite{milc2}. The generation of these configurations was done with the 
canonical anti-periodic temporal boundary conditions for the fermions.

\begin{figure}[t]
\begin{center}
\hspace*{20mm}
\includegraphics*[width=10cm]{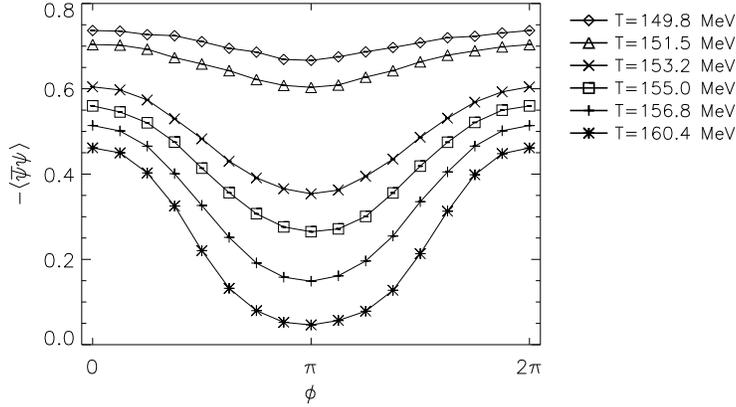}
\end{center}
\caption{\label{fig:chiralcond_dyn}
The bare chiral condensate of the dynamical theory as a function of the boundary
angle $\phi$ for various temperatures.}
\end{figure}

We begin the discussion of the dynamical case with showing in 
Fig.~\ref{fig:chiralcond_dyn} the bare chiral condensate as a function 
of the boundary angle $\phi$. It is obvious, that for the two lowest 
temperatures there is only a small variation with $\phi$, while for 
the larger values of $T$ a much stronger response to changing 
boundary conditions has developed.  

\begin{figure}[b]
\begin{center}
\hspace*{-5mm}\includegraphics*[width=8cm]{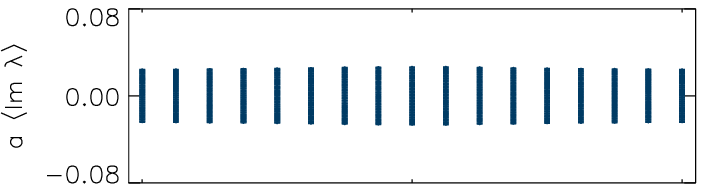}
\vskip3mm
\hspace*{-5mm}\includegraphics*[width=8cm]{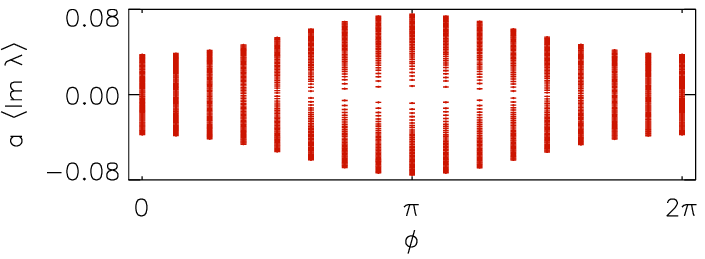}
\end{center}
\caption{\label{fig:gap_dyn}
The 40 lowest eigenvalues in lattice units 
at low ($T = 149.8$ MeV, top plot) and high ($T = 160.4$ MeV, bottom) 
temperatures as a function of the boundary angle $\phi$.}
\end{figure}

In Fig.~\ref{fig:gap_dyn} we study the variation of the 40 lowest eigenvalues 
with the boundary angle $\phi$. We compare the situation in the low temperature
phase ($T = 149.8$ MeV, top plot) and above $T_c$ ($T = 160.4$ MeV, bottom).
A comparison with the quenched case in Fig.~\ref{fig:gap_quench} shows that 
the situation is very similar here: Below $T_c$ the spectrum is almost 
independent of $\phi$, while above $T_c$ the sine-like variation is observed. 
We stress again, that for the unquenched case the Polyakov loop is real, 
since the fermion determinant is much larger for that Polyakov loop sector
(see, e.g., \cite{GaLi}). Thus the shifted scenario depicted in the bottom 
plot of Fig.~\ref{fig:loweval_quench} is absent for the dynamical case.
In this respect the dynamical case is very similar to the behavior of
pure $G_2$ lattice gauge theory \cite{G2}, where the triviality of the center 
also allows only for a sector with real Polyakov loop.

\begin{figure}[t]
\begin{center}
\includegraphics*[width=8cm]{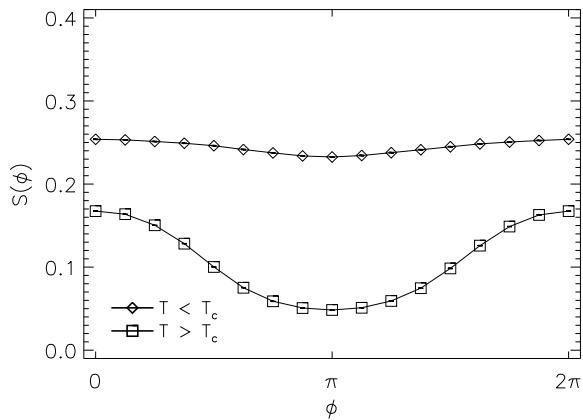}
\end{center}
\caption{\label{fig:integrand_dyn}
The integrand $S(\phi)$ of the chiral condensate for the dynamical case
below and above $T_c$ as a function of the boundary angle $\phi$.}
\end{figure}

The behavior of the IR part of the spectrum shown in Fig.~\ref{fig:gap_dyn}
is inherited by the integrand $S(\phi)$ defined in (\ref{dualconddef2}). In 
Fig.~\ref{fig:integrand_dyn} we show $S(\phi)$ as a function of the 
boundary angle $\phi$ and again compare the situations in the low temperature
phase ($T = 149.8$ MeV) and above $T_c$ ($T = 160.4$ MeV). The 
quark mass $m$ in the definition (\ref{dualconddef2}) of the integrand 
$S(\phi)$ was set to the sea quark mass $m = 0.008$ (in lattice units). 
It is obvious from Fig.~\ref{fig:integrand_dyn}
that the integrand below $T_c$ is rather insensitive to the boundary angle,
while in the high temperature phase we observe a strong variation with 
$\phi$, which in turn gives a larger value for the dual chiral condensate.

\begin{figure}[t]
\begin{center}
\includegraphics*[width=8cm]{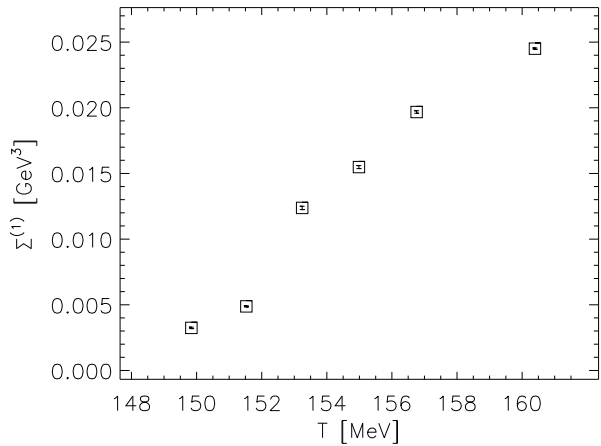}
\end{center}
\caption{\label{fig:dualcond_dyn}
The bare dual chiral condensate $\Sigma^{(1)}$
for the dynamical case as a function 
of the temperature.}
\end{figure}

In Fig.~\ref{fig:dualcond_dyn}, we finally show the dual chiral condensate
$\Sigma^{(1)}$ (without any renormalization) as a function of the temperature. 
Again the quark mass $m$ in the definition (\ref{dualconddef2}) of 
$\Sigma^{(1)}$ was set to the sea quark value $m = 0.008$ (in lattice units).
The plot shows clearly that below $T_c$ the dual chiral condensate is small
but non-vanishing, and above $T_c$ rises quickly with $T$. The observable 
underlines the physical picture of a strongly increased sensitivity to boundary 
conditions above $T_c$.

It is obvious, that the dynamical results presented here have still a 
preliminary character. In particular the temperatures available to us through 
the MILC ensembles come from a rather narrow range around $T_c$. It would be 
interesting how quickly the sensitivity to changing boundary conditions  goes
away for temperatures deeper in the low temperature phase. Also a study with a
chiral  lattice action for the fermions, e.g., the overlap operator would be
desirable,  since this would allow one to cleanly remove the contributions of
the zero modes.
 
\section{Summary, discussion and outlook}

In this contribution we have analyzed the interplay of fermionic temporal 
boundary conditions and spectral quantities of the Dirac operator for 
finite temperature lattice QCD. The temporal boundary conditions are used 
as a tool to probe the system. The underlying working hypothesis is that for 
low temperature, where the temporal extent of the lattice is large -- 
larger than the scale $\Lambda_{QCD}$ of the system -- the spectral 
quantities (and also other observables) are essentially insensitive to a change 
of the boundary conditions. As the temperature $T$ is increased, the temporal 
extent of the lattice shrinks, and near $T_c$ becomes smaller than 
$\Lambda_{QCD}$, and a non-trivial response to 
changing boundary conditions sets in. In particular the following qualitative 
features were demonstrated:

\begin{itemize}

\item
For low temperature the chiral condensate is essentially independent of the
boundary conditions. At high temperature it varies with the boundary angle
$\phi$, and has a minimum for anti-periodic boundary conditions, i.e., $\phi =
\pi$. This observation holds for the dynamical case as well as for quenched 
configurations in the real Polyakov loop sector. For the quenched case with
configurations in the complex Polyakov loop sectors the $\phi$  dependence is
shifted by $\pm 2\pi/3$.  
\vskip3mm

\item 
For low temperatures the spectrum of the Dirac operator is insensitive to 
changing boundary conditions, while above $T_c$ the IR modes show a sine-like 
dependence on the boundary angle $\phi$. For the quenched case, again the 
variation pattern with $\phi$ is shifted by $\pm 2\pi/3$ when ensembles for the
complex Polyakov loop sectors are considered. The sensitivity of the eigenvalues
to changing boundary conditions decreases quickly as one moves towards the UV
end of the spectrum.
\vskip3mm 

\item
The dual chiral condensate is defined as the first Fourier component of the
conventional chiral condensate with respect to the boundary angle $\phi$,  and
thus is an observable testing the sensitivity to the boundary conditions. 
Furthermore, under center rotations (flips) it transforms like the Polyakov loop
and thus is also an order parameter for center symmetry. In the quenched case it
is zero below $T_c$ and non-vanishing above $T_c$. For the fully dynamical case
it  is essentially real and positive, with a small but non-vanishing value below
$T_c$, and a quickly increasing value above the  transition.
\vskip3mm 

\end{itemize}

We stress again that the results for the dynamical case still have a 
preliminary character (see the last paragraph of Section 3). However, the  fact
that also for the full theory the finite temperature transition goes along with
a strong  increase of the sensitivity to temporal fermionic boundary conditions 
is certainly established. 

At this place one may speculate what excitations of the gluon field could drive
the effects observed in the fermionic spectra. Dyons 
(see, e.g., \cite{dyons}) are natural
candidates as their index theorem \cite{Callias} is sensitive to
the boundary conditions: For dyons of given electric and magnetic charge the
zero modes exist only in a certain range of boundary angles. These zero modes then mix and form a near-zero mode band. Dyons can be combined into
calorons \cite{KraanvanBaal,LeeLu}, which are neutral and always possess
zero modes, the latter then hop between the constituent dyons \cite{Perez}.

In a dyon gas model the different dyons appear independently according to their
masses, which in turn are governed by the holonomy. For a holonomy corresponding
to the confined phase all dyons are equally abundant and changing the boundary
conditions will therefore not affect the number of zero modes. For a holonomy
corresponding to the deconfined phase, however, some dyons become heavier than
others and the abundance of zero modes depends on the boundary angle. For SU(2)
the relation to the angle of the Polyakov loop is just as described above
\cite{Luschevskaya}. This suppression effect has been investigated on the
lattice \cite{Martemyanov} and might also explain the decrease of the
topological susceptibility above $T_c$ \cite{Bruckmann}.
These ideas certainly should be substantiated further with lattice methods -- 
a task which we leave for future studies. 

Another interesting and important challenge is to analyze the role of temporal
boundary conditions as a tool for probing finite temperature QCD  using
non-lattice methods.  Only very little has been attempted in this direction so
far  \cite{Jena2,fischer}. Understanding better the response of the  system to
changing boundary conditions certainly will help to characterize the phases of
QCD and the mechanisms driving the transition.

\begin{acknowledgement}
The authors thank Christian Fischer, Tamas Kovacs, Christian Lang, 
Boris Martemyanov, Michael M\"uller-Preussker, Maria Paola
Lombardo, Jan Pawlowski and Andreas Wipf for discussions. This work is partly
supported by the Fonds zur F\"orderung der Wissenschaflichen Forschung under
grants DK W1203-N08,  M1099-N16 and P20330, by the DFG under grant number BR
2872/4-1, and the European Union FP7 program "Hadron Physics 2" grant number 
227431. The simulations were done on
the clusters of the ZID, University of Graz.
\end{acknowledgement}

\end{document}